\documentclass{acm_proc_article-sp}
\usepackage{url}
\begin{document}
\conferenceinfo{EASC}{2015 Edinburgh, UK}
\title{A highly scalable Met Office NERC Cloud model}
\numberofauthors{7}
\author{
\alignauthor Nick Brown\\
\affaddr{EPCC, James Clerk Maxwell Building, Peter Guthrie Tait Road, Edinburgh}\\
\email{nick.brown@ed.ac.uk}
\alignauthor Michele Weiland\\
\affaddr{EPCC, James Clerk Maxwell Building, Peter Guthrie Tait Road, Edinburgh}
\alignauthor Adrian Hill\\
\affaddr{UK Met Office, FitzRoy Road, Exeter, Devon}
\and
\alignauthor Ben Shipway\\
\affaddr{UK Met Office, FitzRoy Road, Exeter, Devon}
\alignauthor Chris Maynard\\
\affaddr{UK Met Office, FitzRoy Road, Exeter, Devon}
\and
\alignauthor Thomas Allen\\
\affaddr{UK Met Office, FitzRoy Road, Exeter, Devon}
\alignauthor Mike Rezny\\
\affaddr{UK Met Office, FitzRoy Road, Exeter, Devon}
}

\maketitle
\begin{abstract}
Large Eddy Simulation is a critical modelling tool for scientists investigating atmospheric flows, turbulence and cloud microphysics. Within the UK, the principal LES model used by the atmospheric research community is the Met Office Large Eddy Model (LEM). The LEM was originally developed in the late 1980s using computational techniques and assumptions of the time, which means that the it does not scale beyond 512 cores. 
In this paper we present the Met Office NERC Cloud model, MONC, which is a re-write of the existing LEM. We discuss the software engineering and architectural decisions made in order to develop a flexible, extensible model which the community can easily customise for their own needs. The scalability of MONC is evaluated, along with numerous additional customisations made to further improve performance at large core counts. The result of this work is a model which delivers to the community significant new scientific modelling capability that takes advantage of the current and future generation HPC machines.
\end{abstract}


\section{Introduction}
Large Eddy Simulation is a computational fluid dynamics technique used to efficiently simulate and study turbulent flows. In atmospheric science, LES are often coupled to cloud microphysics and radiative transfer schemes, to create a high resolution modelling framework that is employed to develop and test physical parametrisations and assumptions used in numerical weather and climate prediction. In the UK, the Met Office Large Eddy Model (LEM) is the principal LES that is used within the Met Office and academia. It includes a detailed cloud microphysics representation and a version of the operational radiative transfer scheme.
The LEM was initially developed in the 1980s and, whilst the scientific output from the model is cutting edge, the code itself has become outdated. Hard coded assumptions made about parallelism, which were sensible 20 years ago, are now the source of severe limitations and this means that the model does not scale beyond 512 cores. This prevents scientists from carrying out very high resolution modelling on the latest HPC machines.

As machines become larger, and significantly different from the architectures that a code was initially designed for, it can sometimes be easier to re-write poorly performing applications rather than attempt to modernise them through re-factoring. The Met Office NERC Cloud model (MONC) is a complete re-write of the LEM, providing the atmospheric scientific community with a tool for modelling atmospheric flows, turbulence and clouds at very high resolutions and/or near real time. The fact that this aspires to be a community code, along with the desire to future proof to as great an extent as possible, has heavily influenced the design of the code. We have adopted a ``plug in" architecture, described in section 3.1, where the model is organised as a series of distinctive, independent, components which can be selected at run-time. This approach, which we discuss in detail, not only allows for a variety of science to be easily integrated but it also supports development targeting different architectures and technologies by simply replacing one component with another.

Other innovative aspects of the code are presented and in particular our approach to data analysis and processing (section 3.2), which is a major feature of the model. These models analyse their raw data to produce higher level information, for instance the average temperature in a cloud or tracking of how specific clouds move through the atmosphere. The existing LEM, like many codes, performs data analysis inline as part of the model timestep which, along with the I/O operation time, is a major bottleneck. In contrast MONC uses the notion of an I/O server, where typically one core per processor is dedicated to handling and analysing the data produced by the model, which is running on the remaining cores. In this manner, MONC can act in a ``fire and forget" fashion, asynchronously sending data to the ``local" I/O server and continuing on with the next timestep whilst it is being processed.

Based upon the innovative approaches adopted, we present in section 4 performance and scalability results of MONC and discuss some of the lessons learnt, both in terms of large scale parallelism and software engineering techniques, that have become apparent in order to reach the level of scalability and performance demanded by the community. Section 5 draws some conclusions and considers future work.

\section{Background}
The Large Eddy Model (LEM) \cite{lem} has been an instrumental tool, used by the weather and climate communities, in modelling clouds and atmospheric flows. Since its inception in the late 1980s, it has been a fundamental tool in the development and testing of the Met Office Unified Model (UM) boundary layer scheme \cite{lock1998}\cite{lock2000}, convection scheme \cite{petch2001}\cite{petch2006} and cloud microphysics \cite{abel2007}\cite{hill2014}. Given the solid scientific basis of the LEM, as established by the inter-comparison studies of \cite{klein2009intercomparison}\cite{ovchinnikov2014intercomparison}\cite{vanzanten2011controls}\cite{fridlind2012comparison}, the community continues to heavily rely on this code as a means for furthering the state of the art. The model was initially developed for single processor, scalar machines and then vectorised to take advantage of the Cray C90, with much of the resulting loop and array structure still present in the current code base. It was not until the mid 1990s when the Met Office took delivery of a Cray T3E, their first parallel machine, that the code was parallelised. Since then, from a software point of view, some perfective maintenance has been performed to allow the LEM to run on later generations of machines but the same basic assumptions and principals have remained unchanged since the code was parallelised or even first written thirty years ago. 

Even for very simple test cases, the LEM scales very poorly beyond 512 cores. A major reason for this is the fact that the LEM only decomposes in one dimension into slices (in X), there are a minimum of two slices per process and due to the way the grid is decomposed, choices for the global domain size in the X direction are severely limited by those in the Y direction. These restrictions place severe limitations upon the model and, whilst one might wish to extend the size in X to gain more parallelism, this also requires an extension in Y which increases the amount of data held locally and often means that the code reaches memory limits. The result is that the community are often forced to run the code unpacked, where all the memory of a node is used but not all the cores, this is a waste of additional compute resource and explicitly required in order to work around the limitations of the current model. Whilst MPI is used for parallelism, the calls are indirect and go via an abstraction layer called GCOM. This is a throwback to the fact that, in the mid 1990s when the model was first parallelised, MPI was not the de-facto standard that it is now and hence it was quite sensible to decouple the communication technology from the actual model. More recently this layer has become more of a hindrance than a help not least because generations of scientists have misunderstood the semantics of the different communication calls. For example, global barriers can often be found intertwined with point-to-point communications without differentiating between memory re-use in buffered and non-blocking sends.

The other important aspect to consider, from a software engineering perspective, is code maintainability. The LEM is written in a mixture of FORTRAN 66, 77 and 90, employing a variety of old fashioned programming constructs such as global variables, gotos and equivalence blocks. This is further exacerbated by the fact that scientists have modified the same files without the enforcement of code standards, so the style is very inconsistent throughout and changes abruptly. Code management is done via a system called nupdate where the code is organised into a snapshot at a specific version called the base, and user code which contains modifications for patching or specific simulations. These files contain, in addition to the code itself, a series of commands such as deleting lines of code from a specific file in the base, modifying existing code or inserting code. These are all fed into nupdate which effectively pre-processes everything into an intermediate, unstructured form which is then compiled. From a user's point of view, one of the major problems is that compiler messages bear no resemblance to their view of the code which can make debugging very difficult to achieve.

This combination of poor scalability, poor performance and antiquated software engineering techniques has meant that the community are now finding it more and more difficult to effectively use this model for the science that they wish to investigate. Modern machines such as ARCHER, a Cray XC30 (the UK national supercomputing service), and the Cray XC40 that the UK Met Office are taking delivery of in 2015 have hundreds of thousands of cores. Many of the problems that the scientific community wish to tackle require parallelism at this level, however the existing LEM can only take advantage of a fraction of the overall capabilities of these machines and as such requires extensive modernisation.

\section{MONC}
In order to support the current and next generation of science we had a choice between refactoring the existing LEM or using the well validated and trusted underlying science of the LEM as the basis for an entirely new model which shares no code. As a result of the common science and scientific assumptions the original LEM can be used for comparison. Because of the many fundamental issues with the LEM, not just in terms of parallelisation but also how the code is written and managed, we elected to follow the re-write avenue. Whilst keeping the same science the re-write route allowed us to use, from day one, modern software engineering and parallelisation techniques. The new model, called the Met Office NERC Cloud model (MONC) is written in Fortran 2003 with MPI for parallelisation and a number of other third party tools, such as Fruit \cite{fruit} for unit testing and Doxygen \cite{doxygen} for documentation. There are two important aims for this code, firstly to provide a community model which is easy and accessible for non HPC experts to modify and extend without having to worry about impacts upon other unrelated areas of the code. Secondly performance and scalability are a major concern for our development of the model and in order to support the scientific community's desired problems the code is firmly targeted at the peta- and exa-scale. 

There is a requirement for the model to support multiple compilers, initially the Cray, GNU, Intel and IBM compilers although this list is subject to change in the future. Whilst compiler implementation of the Fortran 2003 standard has reached maturity in some areas this is not universal and other aspects are not as commonly interpreted or well tested by all. Therefore a unit testing framework, which automatically compiles code and runs the tests using these different technologies is critically important for ensuring specific compiler support and code correctness throughout the development process.

\subsection{Architecture}
MONC has been designed around pluggable components where the majority of the code complexity, including all of the science and parallelisation, are contained within these independent units. They are managed by a registry and at run-time the user selects, via a configuration file, which components to enable. The aim was to make it trivial for a user to add their own components. To encourage this a standard means of definition and interaction with the model has been specified. The majority of a component's functionality is contained within optional callback procedures, which are called by the model at three stages: upon initialization, for each timestep and upon model completion.  There are no global variables in MONC, but instead a user derived type is used to represent the current state of the model and this is passed into each callback which may modify the state. Using this approach means that the model's current state is represented in a structured manner and the type represents a single point of truth about the model's status at any point in time.

\begin{figure*}
\centering
\begin{verbatim}
module test_component
  type(component_descriptor_type) function test_get_descriptor()
    test_get_descriptor%name="test_component"
    test_get_descriptor%version=0.1
    test_get_descriptor%initialisation=>initialisation_callback
    test_get_descriptor%timestep=>timestep_callback
  end function test_get_descriptor

  subroutine initialisation_callback(current_state)
    type(model_state_type), target, intent(inout) :: current_state
    ...
  end subroutine initialisation_callback

  subroutine timestep_callback(current_state)
    type(model_state_type), target, intent(inout) :: current_state
    ...
  end subroutine timestep_callback
end module test_component
\end{verbatim}
\caption{Component standard interface}
\label{fig:component}
\end{figure*}

Figure \ref{fig:component} illustrates the outline of a MONC component, the function \emph{test\_get\_descriptor} provides a descriptor of the component which contains its name, version number and (optionally) populated procedure pointers that represent the callbacks. It can be seen that in this component callbacks have been provided for model initialisation and timestepping. The \emph{initialisation\_callback} and \emph{timestep\_callback} procedures are the actual callbacks themselves and the current model's state is provided via the \emph{current\_state} argument which is of a Fortran derived type and similar to C structs. This \emph{model\_state\_type} derived type contains the current status of the model in a structured manner which the callback procedures may modify. This component is contained within a Fortran module and is picked up by the MONC build system at compile time, and enabled by the user via \emph{test\_component\_enabled=.true.} in the configuration file. The MONC registry, which manages these components, also allows for the user to provide more detailed configuration, for instance, determining the order in which components are run for each different callback. 


Alongside the numerous components representing scientific, parallelism or miscellaneous functionality there is also a model core. This core contains a minimal amount of code to start the model and both manage and support the components themselves. The way in which the core manages components is via a registry, which stores central information about each component and a list of procedure pointers for initialisation, timestepping and finalisation which are called iteratively rather than having to parse each component for every callback. Whilst components are entirely independent from each other and strictly do not interact, it was identified early on in the development process that they often share some common functionality requirements such as the need for logging, data conversions or mathematical functionality. Therefore a series of utilities have been added to the model core, exposed via an API, which provide common functions that components might require and this saves one reinventing the wheel each time a new component is added. 

The model core is mature and the project restrict who may check code in, it is well documented and unit tested to provide a solid foundation for the model. In summary the benefits of adopting a component based architecture for MONC are:

\begin{itemize}
  \item \textit{Trivial to add new components}: Following the standard format these are picked up, included in the model and then simply enabled in the configuration file. Because components are independent and share no code or variables then they simply plug in and out.
  \item \textit{Can add immature components without polluting the rest of the code base}: Due the independent nature of these facets, new functionality can be developed without having to modify existing code. This is important as it allows for additional science to be developed, tested and checked into the code repository without impacting other areas of MONC.
  \item \textit{Simple run-time configuration to customise the model}: A component represents some aspect of the model such as scientific functionality. By adopting this high level approach it is very obvious what functionality is represented in each component and users can easily turn off aspects which are of no interest to their specific run. It is also trivial for users to develop replacement components for areas that they wish to modify or improve. These plug-in via a structure manner. In existing models functionality can often be found a number of levels down in the code, and it can be not only difficult to find calls to disable but also to understand how this might impact the rest of the model.
  \item \textit{Conceptual simplicity}: From a code point of view the running of the model and how each component works via its own callback procedures is a simple concept to understand.
\end{itemize}

The model core also contains an options database, which acts as a centralised store for all model configuration options. When the model is started this database is populated, either from a text configuration file for new simulations or an existing model checkpoint file for continuing simulations. The utilities API of the model core exposes functions to components so that they can check for and retrieve information from this database. Upon a model checkpoint write this centralised store is written to the checkpoint file which allows for simple model restarting.
\subsection{I/O server}
In addition to the simulation itself which produces raw (prognostic) results, lower level data is transformed into higher level (diagnostic) information. This data analysis is a crucial aspect of these models. Traditional approaches inline the data analytical aspect with the rest of a model and run it within in a specific timestep after prognostic data has been generated. However this is not optimal, not just because the data analytics involves significant I/O so the model can be stalled waiting for filesystem access, but also because data analysis work commonly involves intensive communications, for instance when calculating the average values of a global field, and ideally one would overlap this with compute. 

MONC uses an IO server where some of the processes, instead of running the model, are instead dedicated to handling the diagnostic and IO aspects. Typically one core in a processor will run the IO server and this supports the remaining cores running the model. MONC then asynchronously ``fires and forgets'' the raw prognostic data to the IO server for handling. The user configures the IO server via a structured XML configuration file such that the IO server instructs its MONC processes about the specific type of data required and when. Generic actions for handling this data are included with the IO server, which can be added to if required, and are configured in a high level fashion by the user via the IO server XML configuration file. An example of this data analysis to produce two diagnostic outputs; the mean value of a field at each vertical level and secondly the maximum value of a field at each level. The same, \emph{horizontal reduction} action is used by, the first instance configured with the \emph{mean} operator and the second instance configured with the \emph{max} operator. Their high level configuration is all that is required, with the action and underlying framework taking care of the tricky and lower level details such as having to perform inter IO server communications once local values have been computed. The MONC IO server uses a threading approach, where a pool will supply a thread for handling communications from a model process.

There are a number of alternative IO server implementations in use by the community and integration with our own IO server is not mandatory. At the current time of writing, no existing third party IO servers are entirely satisfactory for the diagnostics that the community required from MONC. However, it is important to future proof the model and from the MONC model's point of view it is simply a component, \emph{io\_bridge} which will interface with our IO server. Replacement components, such as \emph{xios\_bridge} can be written to, for instance, interface with the XIOS \cite{xios}  IO server instead. This illustrates an important aspect of the model, where following this pluggable pattern has meant that intricate aspects, such as the handling of diagnostics, is trivial to replace rather than being hard coded in the LEM and other traditional approaches. 
\section{Performance and scaling}
Performance and scalability testing has been conducted with the dry boundary layer test case which models a dry, neutral boundary layer with a constant geostrophic wind. Experiments have been run on the UK national super computing service, ARCHER, a Cray XC30. Each run has modelled 10000 simulation seconds and involves dynamics, pressure solving and the subgrid scheme. The grid is Cartesian, where the size in the vertical (Z) is 64 and that of X and Y is \emph{n$^2$}, where the value of \emph{n} is determined by the desired global size.

\begin{figure}
\includegraphics[scale=0.35]{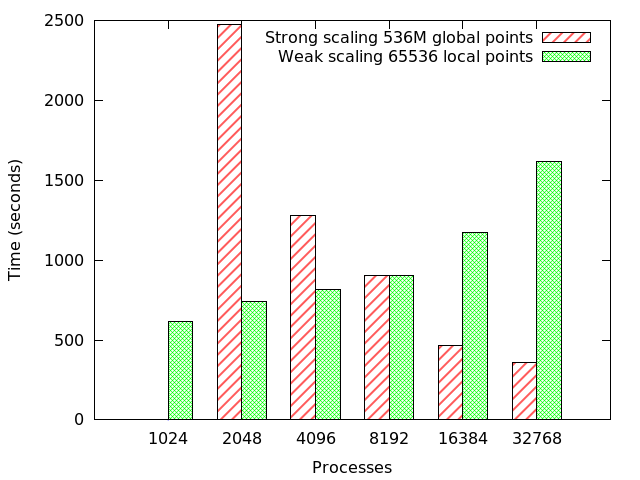}
\caption{MONC scaling experiment}
\label{fig:fftscaling}
\end{figure}

Figure \ref{fig:fftscaling} illustrates MONC scaling. From the strong scaling results it can be seen that, as the number of processes is increased, the run-time for the simulation decreases. However, there is only a small run-time improvement (100 seconds) between running on 16384 and 32768 cores. The weak scaling results, involve 65536 grid points per process (z=64, x=y=32) and provide a clearer picture of the scaling behaviour at larger core counts. The weak scaling run-time results, up until 8096 cores, are reasonably flat however when weak scaling at 16384 cores (global problem size of 1.07 billion global grid points, z=64, x=y=4096) there is a sharp increase in the run-time which is continued at 32768 cores (2.1 billion global grid points, z=64, x=8192, y=4096.) From these results it can be seen that the code, configured in this manner will run at up to 32768 cores and 2.1 billion grid points, although there is some inefficiency which is impacting the run-time as one reaches the larger core counts.

The results presented in figure \ref{fig:fftscaling} use an FFT method for solving pressure terms and analysis at 16k and 32k cores showed that this was taking up a large percentage of the overall run-time. Dealing with pressure terms boils down to solving the Poisson equation and the traditional method involves performing a forward FFT, then in Fourier space solving a vertical ODE before performing a backwards FFT from the spectral domain back to the spatial one. The MONC FFT solver decomposes via a pencil, 2D, decomposition and uses the Fastest Fourier Transformation in the West (FFTW) \cite{fftw} library for the actual FFT computational kernel. However, each FFT requires global all-to-all communications and as one scales up the fact that each process must communicate with every other process for every FFT becomes a bottleneck. 

\begin{figure}
\includegraphics[scale=0.35]{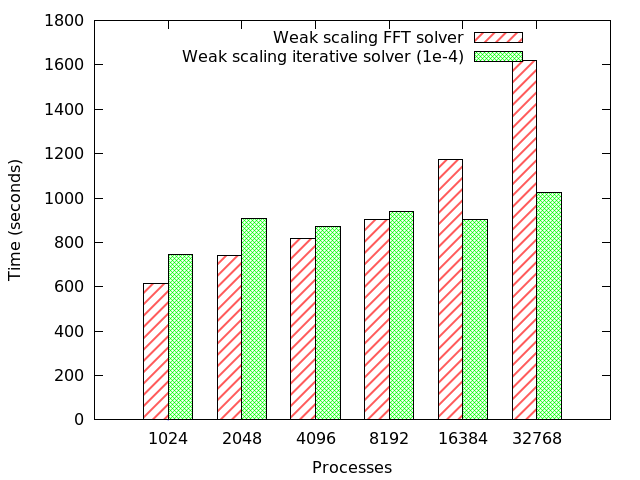}
\caption{FFT vs iterative solver weak scaling}
\label{fig:iterativescaling}
\end{figure}

An iterative solver has also been developed, which solves the Poisson equation using a Krylov subspace method (ILU preconditioned BiCGStab.) The major benefit of this approach is that the only global communication required is a reduction to construct the norm of the residual vector, and all other communications are localised to nearest neighbours for halo swapping. These different solvers have been developed as MONC components, which plug in and out as directed by the user configuration file, and a weak scaling comparison between using an FFT solver and an iterative solver to handle the pressure terms for the dry boundary layer test case are illustrated in figure \ref{fig:iterativescaling}. 

The choice between solvers amounts to a trade off between the lower amount of computation but global all-to-all communication of the FFT solver and more significant amount of computation but less communication of the iterative solver. This can be clearly seen in figure \ref{fig:iterativescaling} where for smaller numbers of cores the FFT solver is more efficient. For instance at 1024 processes solving pressure terms via the FFT solver is 130 seconds faster than using the iterative solver. However as one increases the amount of parallelism this performance gap decreases until the iterative solver overtakes the FFT solver at larger core counts and at 32768 cores the iterative solver reduces the overall run-time by 600 seconds compared to using the FFT solver. The fact that the FFT solver performs so well up until 8096 cores was a surprise to us and this is due to a combination of the very efficient interconnect that can be found on the Cray XC30 along with the highly tuned computation kernels in FFTW.

\begin{figure}
\includegraphics[scale=0.35]{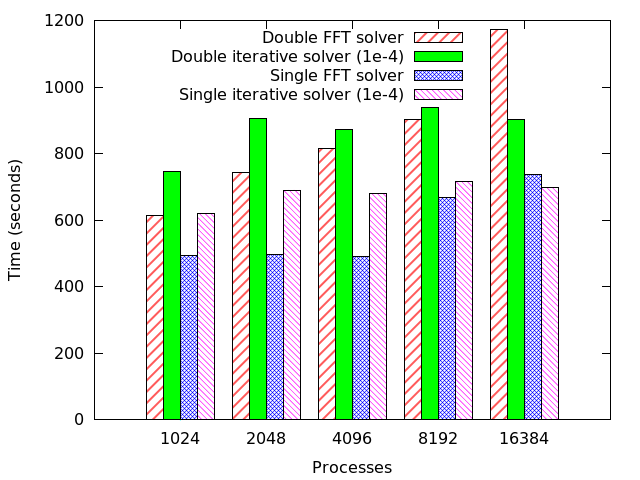}
\caption{Double vs single precision weak scaling}
\label{fig:precisionscaling}
\end{figure}

The results presented so far have all involved the model working in double precision. Whilst some areas of the model must work at this level of precision the pressure solvers do not necessarily need to, especially when solving to 1e-4 which we use in this paper. Instead running the solvers in single precision will not only result in much smaller amounts of data being sent as messages between processes to improve the communication aspects, but will also effectively double the number of elements that can be held in the cache hence improving the computational side of things too. The FFT and iterative solver components were rewritten in single precision, plugged into the model and the weak scaling dry boundary layer test case was rerun on up to 16384 cores. Figure \ref{fig:precisionscaling} illustrates a comparison between the two solvers running at single and double precision for the dry boundary layer test case. It can be seen that single precision provides a performance improvement for both the FFT and iterative solvers but the run-time pattern is similar for single precision as they do for double precision; the FFT solver looks favourable initially and then starts to degrade once the cost of communication becomes significant. At 16384 cores by adopting a single precision iterative solver over the traditional FFT solver for pressure terms, this has resulted in an run-time reduction of 476 seconds. It can be clearly seen that single precision, if a weak stopping criteria can be tolerated, does make a difference and is an important optimisation that can be easily applied with our plugable component architecture. 

\section{Conclusions and further work}
This paper has described the MONC model, from a software engineering and architectural point of view, which delivers a step change in scalability, performance and capability compared to the existing LEM model. We have described the component based architecture and discussed how this forms the basis for a flexible and extensible code base which the community can easily add their own science to. This simple conceptual view of the model allows the user to easily configure MONC for their own requirements and ensures that run-time is not being wasted in areas not required for a specific simulation. Crucial to performance is how one handles the data analysis aspects of the model and our approach, using an IO server approach to effectively separate this from the raw science, has been introduced. We have demonstrated scalability up to 32678 cores and discussed some of the crucial factors that impact performance at this core count and how the architecture of the model is suited for allowing users to trivially experiment with these aspects. 

As the scientific community start to pick up this new model, add their own components and use it in their research, there is still further work to be done from a software point of view. Based upon the results in this paper, it will be interesting to further investigate some of the techniques which have given performance improvements. A bespoke preconditioner, which exploits the problem's known mathematical structure, can be developed which boosts performance of the solver compared to the generic ILU preconditioner. If greater accuracy is required then a mixed-precision solver can be developed, which exploits single precision to achieve performance with a restart in double precision to achieve the desired accuracy. An additional benefit of such a solver is that performance tuning it being done dynamically by the model, rather than relying upon the user. The component based architecture also lends itself to providing support for the model on other platforms, for instance, by developing a number of GPU based components to take advantage of these machines. 
\bibliographystyle{plain}
\bibliography{monc-easc}

\end{document}